\documentclass[9pt,twocolumn,twoside]{osajnl}

\journal{ol} 

\setboolean{shortarticle}{true} 

\ifthenelse{\boolean{shortarticle}}{\colorlet{color2}{color2b}}{\colorlet{color2}{color2}} 

\title{Degenerate cavity supporting more than 31 Laguerre-Gaussian modes}

\author[1,2]{Ze-Di Cheng}
\author[1,2]{Zhao-Di Liu}
\author[1,2]{Xi-Wang Luo}
\author[1,2]{Zheng-Wei Zhou}
\author[1,2]{Jian Wang}
\author[1,2]{Qiang Li}
\author[1,2]{Yi-Tao Wang}
\author[1,2]{Jian-Shun Tang}
\author[1,2*]{Jin-Shi Xu}
\author[1,2\dag]{Chuan-Feng Li}
\author[1,2]{Guang-Can Guo}

\affil[1]{CAS Key Laboratory of Quantum Information, University of Science and Technology of China, Hefei 230026, People's Republic of China}
\affil[2]{Synergetic Innovation Center of Quantum Information and Quantum Physics, University of Science and Technology of China, Hefei 230026, People's Republic of China}

\affil[*]{Corresponding author: jsxu@ustc.edu.cn}
\affil[$\dag$]{Corresponding author: cfli@ustc.edu.cn}

\dates{Compiled \today}

\ociscodes{(050.4865) Optical vortices; (230.5750) Resonators; (000.2190) Experimental physics.}

\doi{\url{http://dx.doi.org/10.1364/ol.XX.XXXXXX}}

\begin{abstract}
Photons propagating in Laguerre-Gaussian modes have characteristic orbital angular momenta, which are fundamental optical degrees of freedom. The orbital angular momentum of light has potential application in high capacity optical communication and even in quantum information processing. In this work, we experimentally construct a ring cavity with 4 lenses and 4 mirrors that is completely degenerate for Laguerre-Gaussian modes. By measuring the transmitted peaks and patterns of different modes, the ring cavity is shown to supporting more than 31 Laguerre-Gaussian modes. The constructed degenerate cavity opens a new way for using the unlimited resource of available angular momentum states simultaneously.
\end{abstract}

\setboolean{displaycopyright}{true}

\begin{document}

\maketitle
\thispagestyle{fancy}
\ifthenelse{\boolean{shortarticle}}{\abscontent}{}

The Laguerre-Gaussian (LG) modes are solutions for beam profiles with circularly symmetric. They are written in cylindrical coordinates using Laguerre polynomials and shown to obtained well-defined orbital angular momentum (OAM) \cite{OAM}. The phase fronts of light beams in OAM eigenstates rotate, clockwise for positive OAM values, anti-clockwise for negative values, which could result in some unique features. The synthetic dimension with the values of OAM defined as the dimensional basis is recognized as a unique asset in many studies,  including high-capacity optical communication \cite{usage4,usage6}, versatile optical tweezers \cite{tweezer}, quantum information and quantum foundation \cite{eng3,eng5,eng6}. It has been recently demonstrated quantum entanglement involving angular momenta as high as hundreds \cite{eng5,eng2}. The manipulation and measurement of OAM states can be reached with high precision \cite{tec2,eng3,tec4,tec5}, which leads to new applications of OAM states, such as detecting of a spinning object and lateral motion \cite{new1,new2}.

Although, in principle, there are infinite synthetic degrees of freedom for OAM states, how to construct some functional devices by manipulating them in the synthetic dimension become a subject to be developed. Recently, a new kind of photonics simulator based on energy-degenerated optical cavities has been theoretically presented, which can support a large number of OAM modes \cite{quasim}. By taking advantage of such kind of optical cavity, one can simulate photonics topological matters, as well as construct all-optical devices by manipulating in photonic synthetic dimension \cite{xiwangluo,luo2016synthetic}. All such potentially novel applications are ascribed to set up a kind of degenerate optical cavity, which can support plenty of energy-degenerated OAM modes.

Here, we present an experimental framework using numbers of OAM states in an optical cavity simultaneously, which is referred to a degenerate cavity. Different from previous works with OAM in cavities \cite{cav_image,laser_oam}, we precisely measure the transmitted peaks and beam profles of different modes in the cavity, which is shown to supporting more than 31 LG modes. Moreover, the constructed cavity is in principle completely degenerate, which can be used to generate lasers and entangled photon sources with high dimensional LG modes \cite{referee1,referee2,referee3,referee4} and other related fields \cite{mulitymode,natsyn,ncref2}.

The theoretical framework of a degenerate cavity was introduced in \cite{degeneratei,degenerateii}. From a geometrical optics point of view, an optical cavity is degenerate when an arbitrary ray retraces its own path after a single round trip. To understand the design principles of completely degenerate cavities, we consider the propagation of light field in a cavity between two planes perpendicular to the axis of the cavity. The transmitted electric field $E_{1}(x_{1},y_{1})$ can be written in the form of the Collins integral \cite{Collins:70}
\begin{equation}
\begin{split}
e^{-ikz_{1}}E_{1}(x_{1},y_{1})=e^{-ikL}e^{-ikz_{0}} \frac{i}{ \lambda B}\int\!\!\!\!\int E_{0}(x_{0},y_{0})\times \\
e^{[- \frac{i}{\lambda B}(Ax_{0}^{2}+Dx_{1}^{2}-2x_{0}x_{1}+Ay_{0}^{2}+Dy_{1}^{2}-2y_{0}y_{1})]}dx_{0}dy_{0} ,
\end{split}
\end{equation}
where $\lambda$ and $k$ are the wavelength and wave number. $E_{0}(x_{0},y_{0})$ is the initial electric transverse field. $L$ is the length of the optical path along the optical axis between the two planes. The transmitted and initial fields are connected through the ray matrix between the two planes with elements $A, B, C$ and $D$ \cite{laser}.

\begin{figure}[tb]
\centering
\includegraphics[width=\linewidth]{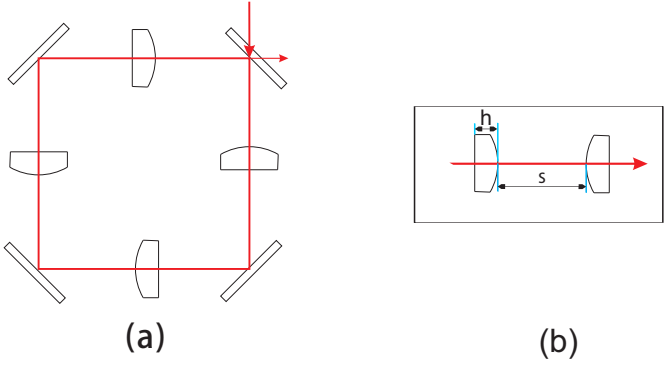}
\caption{Schematic diagram of optical design. (a) A planar graph of our cavity. (b) The implementation of a $4f$ lens system in practice. s is the distance  between the two lenses, and h is the thickness of the lens.}
\label{Fig1}
\end{figure}

\begin{figure}[tb]
\centering
\includegraphics[width=\linewidth]{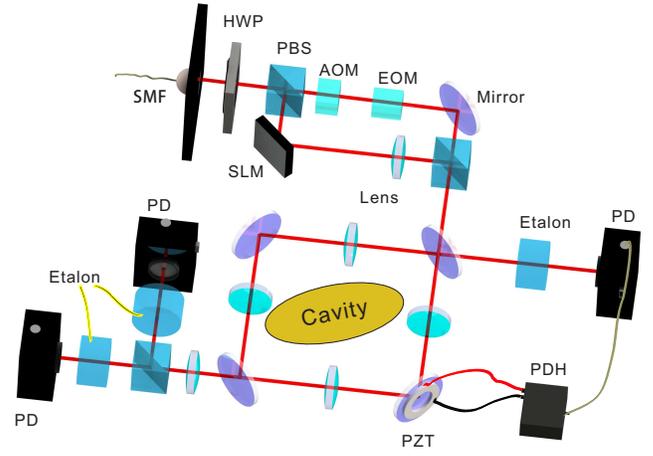}
\caption{
Experimental setup. The degenerate cavity consists of 4 lenses and 4 mirrors. The laser from a single mode fiber (SMF) is split into two parts by a polarization beam splitter (PBS). One is used as a reference light and to lock the cavity. The other is used as the signal. The half wave plate (HWP) is used to adjust the intensity ratio of the split beams. The acoustic optical modulator (AOM) is used to shift the frequency of the reference light. The electro-optical modulator (EOM) is used to do sideband modulation to lock the cavity. The signal beam illuminated on the screen of a spatial light modulator (SLM) is transformed into LG modes. Then these two beams are combined by a PBS and pumped into the cavity. An etalon is used to filter out the reference laser from the beam reflected by the cavity mirror and the reference light is detected by a photoelectric detector (PD). The electrical signal is leaded to an equipment using Pound-Drever-Hall (PDH) technique to lock the cavity by driving the piezoelectric transducer (PZT). Another two etalons are used to distinguish the reference laser and signal laser from the transmission beam, which are separated into two beams and detected by PDs.}
\label{Fig2}
\end{figure}

\begin{figure}[tb]
\centering
\includegraphics[width=\linewidth]{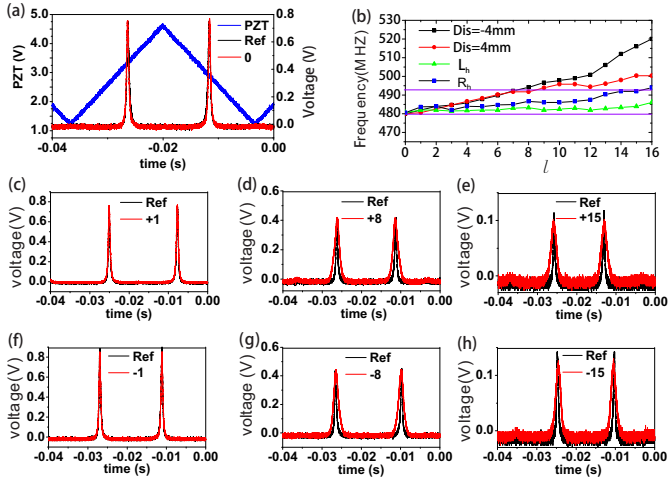}
\caption{
The transmission peaks of the modes. (a) The Gaussian mode. The blue triangular wave corresponds to the electric signal driving the piezoelectric transducer (PZT), and the peaks correspond to transmission peaks of reference light (Ref, black line) and signal beam (red line labelled with the topological charge $l$) respectively. The left coordinate scale corresponds to the signal driving the PZT while the right coordinate scale corresponds to the transmission peaks. (b) The dispersion relationship of our cavity. The purple lines define the linewidth of the cavity. $\rm R_{h}$ and $\rm L_{h}$ correspond to right-handed and left-handed rotating vortex beams in degenerate situation, while Dis represents that one of the 4 lenses subjects to some displacements from the degenerate situation. (c)-(e) The transmission peaks of left-handed rotating vortex beams in degenerate situation. (f)-(h) The transmission peaks of right-handed rotating vortex beams.}
\label{Fig3}
\end{figure}

\begin{figure}[!h]
\centering
\includegraphics[width=\linewidth]{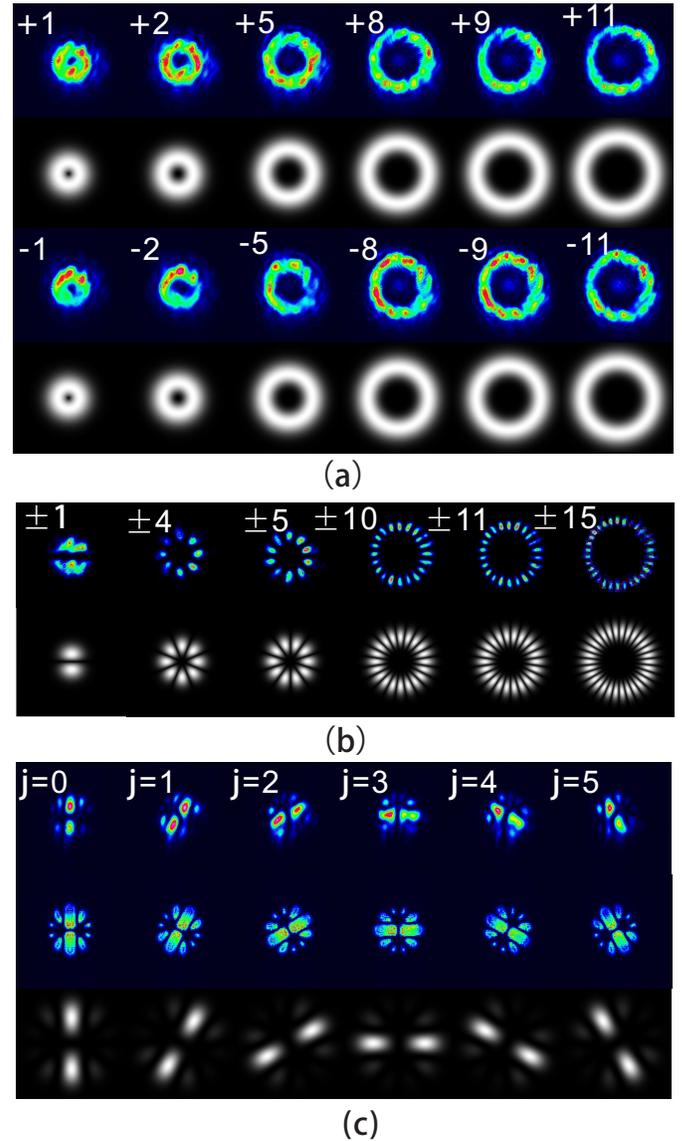}
\caption{
Experimental results of beam profiles. (a) The transmission beam profiles of LG modes with topological charges signed in the corners. The black-white images are theoretical results while the coloured pictures are experimental results. (b) The transmission beam profiles of conjugate superposition states with topological charges signed in the corners. (c) The transmission beam profiles of a class of six-dimension orthographical superposition states corresponding to $|\varPhi_{\rm{j}}\rangle$. The profiles in the second row are experimental results before the cavity.}
\label{Fig4}
\end{figure}

The resonate frequencies and eigenmodes of the cavity can be solved by using the condition that the field must reproduce itself after a round trip in the cavity. If the optical elements have cylindrical symmetry, the solutions happen to be the LG modes $E_{p,l}e^{-ikz}$ with the transverse field \cite{laser}
\begin{equation}
\begin{split}
E_{p,l}(\rho,\varphi)=& E_{0} \sqrt{\frac{2p!}{\pi(|l|+p)!}} \frac{1}{w(z)} (\frac{\sqrt{2}\rho}{w(z)})^{|l|}\times\\
&L_{p}^{|l|}(\frac{2\rho^{2}}{w^{2}(z)}) e^{\frac{-\rho^{2}}{w^{2}(z)}}e^{\frac{-ik\rho^{2}}{2R(z)}}e^{-i[2p+|l|+1]\psi(z)}e^{-il\varphi},
\end{split}
\end{equation} 
where $\rho, \varphi$ and $z$ are the parameters of polar coordinate system. $p$ and $l$ are the parameters of the Laguerre polynomials and $l$ identifies the topological charge of the corresponding vortex beam. $w(z)$ is the transverse width of the light beam. $R(z)$ is the wavefront curvature radius. $\psi(z)$ is the Gouy phase, and $L_{p}^{|l|}(x)$ is the generalized Laguerre polynomial.

The resonate frequency for each $E_{p,l}$ mode in a ring-type cavity is determined by \cite{degeneratei}
\begin{equation}
kL_{0}-(2p+l+1)arccos \frac{A+D}{2}=2n\pi,
\end{equation}
where $n$ is an integer, $L_{0}$ is the length of the round-trip optical path. When $A+D=2$, the resonate frequency is independent of parameters $p$ and $l$, which means the cavity is degenerate for spatial modes. The off diagonal elements of the round-trip ray matrix, $B$ and $C$, only affect the beam waist $w_{0}$ of the resonate modes.

We consider a cavity consisting of 4 lenses arranging in a 4f lens system just as shown in Fig.~\ref{Fig1} (a). This kind of cavity has been shown to satify the degenerate condition and the corresponding matrix happens to be the unitary matrix ($A=D=1, B=C=0$) \cite{laser}. However, in practical, the distance between the surface of two lenses are not $2f$. Considering a ray of light pass through a planoconvex lens as shown in Fig.~\ref{Fig1} (b) and the ray matrix can be written as
\begin{equation}
M_{r}=\left[ \begin{matrix}
1&0\\
0& \frac{n_{2}}{n_{1}}
\end{matrix}
 \right]\left[ \begin{matrix}
1&h\\
0&1
\end{matrix}
 \right]\left[ \begin{matrix}
1&0\\
\frac{n_{1}-n_{2}}{n_{2}r}& \frac{n_{1}}{n_{2}}
\end{matrix}
 \right],
\end{equation} 
where $n_{1}$ and  $n_{2}$ are the refractive index of free space and lens respectively, $r$ is the radius of curvature of the lens and $h$ is the   thickness of the lens. A physical lens can be equivalent to an ideal lens with focal length $f$ depending on parameters of the lens and a free space of length $L_{eq}$, namely
\begin{equation}
M_{r}=\left[ \begin{matrix}
1&L_{eq}\\
0&1
\end{matrix}
 \right]\left[ \begin{matrix}
1&0\\
-\frac{1}{f}&1
\end{matrix}
 \right].
\end{equation}
As a result, the distance between two adjacent lens in the degenerate cavity should be
$s=2f-L_{eq}.$ Finally, we get the free spectral range (FSR) of our cavity $FSR=\frac{c}{4s+4n_{2}h},$ where $c$ is the velocity of light. To make it clear, 4 planoconvex lenses are used in the cavity, the center thickness of the lens is approximating 4.0 mm, the refractive index is 1.51, the radius of the curvature is about 38.9 mm, and the clear aperture is about 22.8 mm. The size of the beam on the lens is about 6.0 mm for LG mode with charge number 15. Therefore, we get $FSR\approx480MHz$. The reflectivity of two mirrors are $99.99\%$ and other two are 93\%. While the transmittance of the four lenses are 99.9\%, so the linewidth of the cavity is calculated to be about $12MHz$.

The experimental setup is shown in Fig.~\ref{Fig2}. A continuos ultra-narrow bandwidth laser (MBR, Coherent Int.) is splitted into two beams. One is used as a reference light and to lock the cavity. The other is used as the signal. The reference light is shifted about one FSR of our cavity via an acoustic optical modulator (AOM). The electro-optical modulator (EOM) is used to assist on locking the cavity in which the Pound-Drever-Hall (PDH) technique is applied. The signal beam illuminated on the screen of a spatial light modulator (SLM) loaded with hologram \cite{tec2} is transformed into LG modes. Then these two beams are combined and pumped into the cavity. Three etalons with linewidth approximating $100$ MHz and FSR approximating $7$ GHz are used to distinguish the reference laser and signal laser. We then use photoelectric detectors (PDs, PDA10A-EC, Thorlabs Inc.) to record the intensity of the output light beams. If we want to detect the beam profiles, the detector is replaced by a camera (BC106N-VIS, Thorlabs Inc., not shown in Fig.~\ref{Fig2}). 

In the experiment, we fix the frequency of the microwave driving the AOM and then drive the piezoelectric transducer (PZT) attached to one of the cavity mirrors to scan the transmitted peaks. In Fig.~\ref{Fig3} (a), the triangular wave corresponds to the electric signal driving the PZT, which changes the cavity length accordingly. The transmitted peak signals are then recorded. The black line corresponds to the signal of reference light and the red line presents the signal of Gaussian mode ($l$=0), which almost completely overlap with each other. Fig.~\ref{Fig3} (c)-(e) show the left-handed rotating vortex beams and Fig.~\ref{Fig3} (f)-(h) show the right-handed rotating vortex beams, with the corresponding topological charges signed above the peaks. From the figures we can see that the transmitted peaks of LG modes coincide with the reference light roughly, which means different LG modes resonate at the same point and the cavity is degenerate.

Fig.~\ref{Fig3} (b) shows the dispersion relationship of the cavity. As we can see in Fig.~\ref{Fig3} (a), even the two peaks coincide with each other mostly, there does exist differences. The frequency driving the AOM is fine tuned to make the peaks coinciding perfectly, and the corresponding frequencies are recorded. Horizontal axis is the topological charge of the pumped beam, and longitudinal axis shows the frequency differences between reference light and signal light. The purple lines separated by $12$ MHz corresponds to the linewidth of the cavity. $\rm R_{h}$ and $\rm L_{h}$ correspond to right-handed and left-handed rotating vortex beams in degenerate situation, while Dis represents the right-handed signals that one of the 4 lenses leads to some displacements from the degenerate situation. From the figure we can see that in the degenerate situation the differences between modes with topological charges from 0 to 15 are limited to $12$MHz. However, in the displaced situations, the differences increase almost linearly. When the topological charge is only 7, it has already surpassed the region of the cavity linewidth.  The frequency of $\rm L_{h}$ increase more slowly than that of $\rm R_{h}$. When the topological charge equals to 16, $L_{h}$ is still well within the linewidth of the cavity but $R_h$ locates at the margin of the linewidth of the cavity.

Next, we fix the frequency of the microwave driving the AOM and lock the cavity to the reference light. Fig.~\ref{Fig4} shows the transmitted beam profiles with different LG modes. The coloured ones are experimental results and the black-white ones are the corresponding theoretical predictions. From Fig.~\ref{Fig4} (a), we can see the eigen shapes of vortex beams clearly with topological charges signed in the corners. Fig.~\ref{Fig4} (b) shows beam profiles of conjugate superposition states, namely
\begin{equation}
|\Psi\rangle=\frac{1}{\sqrt{2}}\left(|+l\rangle+|-l\rangle\right).
\end{equation}
The experimental beam profiles with petals agree well with the theoretical predictions. We further prepare a class of six-dimensional orthographical superposition states given by
\begin{equation}
\left(
\begin{matrix}
\varPhi_{0}\\
\varPhi_{1}\\
\varPhi_{2}\\
\varPhi_{3}\\
\varPhi_{4}\\
\varPhi_{5}
\end{matrix}
  \right)
=\frac{1}{\sqrt{6}}\left(
 \begin{matrix}
  1&1&1&1&1&1\\
  1&\omega&\omega^2&\omega^3&\omega^4&\omega^5\\
  1&\omega^2&\omega^4&\omega^6&\omega^8&\omega^{10}\\
  1&\omega^3&\omega^6&\omega^9&\omega^{12}&\omega^{15}\\
  1&\omega^4&\omega^8&\omega^{12}&\omega^{16}&\omega^{20}\\
  1&\omega^5&\omega^{10}&\omega^{15}&\omega^{20}&\omega^{25}
  \end{matrix}
  \right)\left(
  \begin{matrix}
  |+1\rangle\\
  |-1\rangle\\
  |+3\rangle\\
  |-3\rangle\\
  |+5\rangle\\
  |-5\rangle
  \end{matrix}
  \right)
\end{equation}
with $\omega=e^{i\pi/3}$. The beam profiles in the first and second rows in Fig.~\ref{Fig4}c represent the corresponding experimental results after and before the cavity, respectively. They agree with each other very well, and both coincide perfectly with the theoretical one. As a result, the cavity maintains well the beam profiles.

In summary, we have constructed a degenerate ring cavity supporting more than 31 LG modes simultaneously. However, due to the imperfect optical elements in the cavity, especially the radially varying thickness of the lenses, the resonate frequencies of LG modes with different charge numbers deviate from each other. Moreover, the imperfect arrangement of the optical elements which is strictly required by a completely degenerate cavity would further lead to an non-unitary transmission matrix. Even for the light beams with the same absolute topological charges ($|l|$), the resonant frequencies are shown to be different (Fig.~\ref{Fig4}b), which is due to fact that LG modes generated through the SLM are not ideal.

Nevertheless, our work represents the first experimental work demonstrating the degenerate cavity supporting many LG modes. Since the LG modes form a complete basis, the constructed cavity can be used to supporting other kinds of beam structures. As a completely degenerate cavity in principle, it can also be used to generate lasers and entangled photon sources with high dimensional LG modes \cite{referee1,referee2,referee3,referee4} and other related fields \cite{mulitymode,natsyn,ncref2}. Our work opens the way to manipulate the synthetic degrees of freedom for OAM states.

\rule{0pt}{0pt} 

\noindent{\bf Funding.} This work was supported by the National Key Research and Development Program of China (Grants No. 2016YFA0302700), the National Natural Science Foundation of China (Grants No. 61327901, 11325419, 11274297, 61322506 and 61490711 ), the Strategic Priority Research Program (B) of CAS (Grant No. XDB01030300), the Key Research Program of Frontier Sciences, CAS (No. QYZDY-SSW-SLH003), the Fundamental Research Funds for the Central Universities (Grants No. WK2470000020), and the Youth Innovation Promotion Association and Excellent Young Scientist Program CAS.


\begin{thebibliography}{10}
\newcommand{\enquote}[1]{``#1''}

\bibitem{OAM}
L.~Allen, M.~W. Beijersbergen, R.~J.~C. Spreeuw, and J.~P. Woerdman, Phys. Rev.
  A \textbf{45}, 8185 (1992).

\bibitem{usage4}
I.~B. Djordjevic and M.~Arabaci, Opt. Express \textbf{18}, 24722 (2010).

\bibitem{usage6}
Y.~Yan, G.~Xie, M.~P. Lavery, H.~Huang, N.~Ahmed, C.~Bao, Y.~Ren, Y.~Cao,
  L.~Li, Z.~Zhao, A.~F. Molisch, M.~Tur, M.~J. Padgett, and A.~E. Willner, Nat.
  Commun. \textbf{5}, 4876 (2014).

\bibitem{tweezer}
M.~Padgett and R.~Bowman, Nat. Photon. \textbf{5}, 343 (2011).

\bibitem{eng3}
A.~Mair, A.~Vaziri, G.~Weihs, and A.~Zeilinger, Nature \textbf{412}, 313
  (2001).

\bibitem{eng5}
R.~Fickler, R.~Lapkiewicz, W.~N. Plick, M.~Krenn, C.~Schaeff, S.~Ramelow, and
  A.~Zeilinger, Science \textbf{338}, 640 (2012).

\bibitem{eng6}
Z.-Q. Zhou, Y.-L. Hua, X.~Liu, G.~Chen, J.-S. Xu, Y.-J. Han, C.-F. Li, and
  G.-C. Guo, Phys. Rev. Lett. \textbf{115}, 070502 (2015).

\bibitem{eng2}
M.~Krenn, M.~Huber, R.~Fickler, R.~Lapkiewicz, S.~Ramelow, and A.~Zeilinger, P.
  Natl. Acad. Sci. \textbf{111}, 6243 (2014).

\bibitem{tec2}
E.~Bolduc, N.~Bent, E.~Santamato, E.~Karimi, and R.~W. Boyd, Opt. Lett.
  \textbf{38}, 3546 (2013).

\bibitem{tec4}
W.~Lee, X.-C. Yuan, and W.~Cheong, Opt. Lett. \textbf{29}, 1796 (2004).

\bibitem{tec5}
M.~Malik, M.~Mirhosseini, M.~P. Lavery, J.~Leach, M.~J. Padgett, and R.~W.
  Boyd, Nat. Commun. \textbf{5}, 3115 (2014).

\bibitem{new1}
M.~P. Lavery, F.~C. Speirits, S.~M. Barnett, and M.~J. Padgett, Science
  \textbf{341}, 537 (2013).

\bibitem{new2}
N.~Cvijetic, G.~Milione, E.~Ip, and T.~Wang, Sci. Rep. \textbf{5}, 15422
  (2015).

\bibitem{quasim}
X.-W. Luo, X.~Zhou, C.-F. Li, J.-S. Xu, G.-C. Guo, and Z.-W. Zhou, Nat. Commun.
  \textbf{6}, 7704 (2015).

\bibitem{xiwangluo}
X.-F. Zhou, X.-W. Luo, S.~Wang, G.-C. Guo, X.~Zhou, H.~Pu, and Z.-W. Zhou,
  Phys. Rev. Lett. \textbf{118}, 083603 (2017).

\bibitem{luo2016synthetic}
X.-W. Luo, X.~Zhou, J.-S. Xu, C.-F. Li, G.-C. Guo, C.~Zhang, and Z.-W. Zhou,
  arXiv:1612.08467  (2016).

\bibitem{cav_image}
S.~Gigan, L.~Lopez, N.~Treps, A.~Ma{\^\i}tre, and C.~Fabre, Phys. Rev. A
  \textbf{72}, 023804 (2005).

\bibitem{laser_oam}
D.~Naidoo, K.~A{\"\i}t-Ameur, and A.~Forbes, \enquote{Intracavity vortex beam
  generation,} in \enquote{SPIE Optical Engineering+ Applications,}
  (International Society for Optics and Photonics, 2011), pp. 813009--813009.

\bibitem{referee1}
M.~Martinelli, J.~A.~O. Huguenin, P.~Nussenzveig, and A.~Z. Khoury, Phys. Rev.
  A \textbf{70}, 013812 (2004).

\bibitem{referee2}
B.~C. dos Santos, A.~Z. Khoury, and J.~A.~O. Huguenin, Opt. Lett. \textbf{33},
  2803 (2008).

\bibitem{referee3}
B.~C. dos Santos, K.~Dechoum, and A.~Z. Khoury, Phys. Rev. Lett. \textbf{103},
  230503 (2009).

\bibitem{referee4}
K.~Liu, J.~Guo, C.~Cai, S.~Guo, and J.~Gao, Phys. Rev. Lett. \textbf{113},
  170501 (2014).

\bibitem{mulitymode}
A.~J. Koll{\'a}r, A.~T. Papageorge, V.~D. Vaidya, Y.~Guo, J.~Keeling, and B.~L.
  Lev, Nat. Commun. \textbf{8}, 14386 (2017).

\bibitem{natsyn}
N.~Schine, A.~Ryou, A.~Gromov, A.~Sommer, and J.~Simon, Nature \textbf{534},
  671 (2016).

\bibitem{ncref2}
S.~Ngcobo, I.~Litvin, L.~Burger, and A.~Forbes, Nat. Commun. \textbf{4}, 2289
  (2013).

\bibitem{degeneratei}
J.~A. Arnaud, Appl. Opt. \textbf{8}, 189 (1969).

\bibitem{degenerateii}
J.~A. Arnaud, Appl. Opt. \textbf{8}, 1909 (1969).

\bibitem{Collins:70}
S.~A. Collins, J. Opt. Soc. Am. \textbf{60}, 1168 (1970).

\bibitem{laser}
N.~Hodgson and H.~Weber, \emph{Laser Resonators and Beam Propagation:
  Fundamentals, Advanced Concepts, Applications}, vol. 108 (Springer, 2005).

\end{thebibliography}
%

\end{document}